\documentclass[aps, twocolumn,prl,showpacs,superscriptaddress]{revtex4}

\usepackage{graphicx}
\usepackage{dcolumn}
\usepackage{bm}

\begin{document}

\title{Rapid diffusion of dipolar order enhances dynamic nuclear polarization}\

\author{Anatoly E. Dementyev}
\affiliation{Francis Bitter Magnet Laboratory,}
\author{David G. Cory}
\affiliation{Francis Bitter Magnet Laboratory,}
\affiliation{Department of Nuclear Science and Engineering, Massachusetts Institute of Technology, Cambridge, MA 02139, USA}
\author{Chandrasekhar Ramanathan\footnote{Author to whom correspondence should be addressed. Electronic address:
sekhar@mit.edu}}
\affiliation{Francis Bitter Magnet Laboratory,}
\affiliation{Department of Nuclear Science and Engineering, Massachusetts Institute of Technology, Cambridge, MA 02139, USA}

\date{\today}

\begin{abstract}

In a dynamic nuclear polarization experiment on a 40 mM solution of 4-amino-TEMPO in a 40:60 water/glycerol mixture, we have observed that the  bulk dipolar reservoir is cooled to a spin temperature of 15.5 $\mu$K, following microwave irradiation for 800 s.  This is significantly cooler than the 35 mK spin temperature of the Zeeman reservoir.  Equilibration of the two reservoirs results in a 50 \% increase in the NMR signal intensity, corresponding to a Zeeman spin temperature of 23 mK.  In order to achieve this polarization directly, it was necessary to irradiate the sample with microwaves for 1500 s.  Cooling of the dipolar reservoir occurs during polarization transport through the magnetic field gradient around the paramagnetic impurity, and is rapidly communicated to the bulk by dipolar spin diffusion.  As dipolar spin diffusion is significantly faster than Zeeman spin diffusion, the bulk dipolar reservoir cools faster than the Zeeman reservoir.   This process can be exploited to rapidly polarize the nuclear spins, by repeatedly cooling the dipolar system and transferring the polarization to the Zeeman reservoir.

\end{abstract}

\pacs{ }

\maketitle

Dynamic nuclear polarization (DNP) has been extensively used to study ordering in dielectric crystals, for the creation of polarized spin targets, and for NMR signal enhancement \cite{Abragam-1982}.  Microwave irradiation of a coupled electron-nuclear spin system can facilitate a transfer of polarization from the electron to the nuclear spin.  In the case of dielectric materials considered here, the electron spins are localized and the non-equilibrium polarization of the bulk nuclei is generated via a two-stage process:  a polarization exchange local to the defect; and spin transport to distribute the polarization throughout the sample.   Here the DNP process is essentially the inverse of the standard $T_1$ relaxation mechanism in dielectric solids \cite{Bloembergen-1949}.  

In dielectric materials DNP typically occurs via either the solid effect or thermal mixing \cite{Abragam-1982,Farrar-2001,Reynhardt-2003b}.  The solid effect occurs in systems in which the electron-nuclear coupling is not purely isotropic.  The anisotropic coupling mixes the nuclear Zeeman levels, and the forbidden transitions become partially allowed.  High power microwaves are used to saturate one of the forbidden transitions, and rapid electron spin relaxation creates the non-equilibrium nuclear polarization.  Thermal mixing occurs in systems in which the linewidth of the electron spins is larger than the nuclear Zeeman frequency.  If the linewidth is homogeneous, off-resonance microwave irradiation of the electron spin line produces a cooling of the electron spin dipolar reservoir.  A flip-flop interaction between two electron spins in the same line (separated by the nuclear Zeeman frequency), flips the nuclear spin. If the electron spin linewidth is inhomogeneous, thermal mixing occurs via cross relaxation between different electron spin packets.  

\begin{figure}
\scalebox{0.4}{\includegraphics{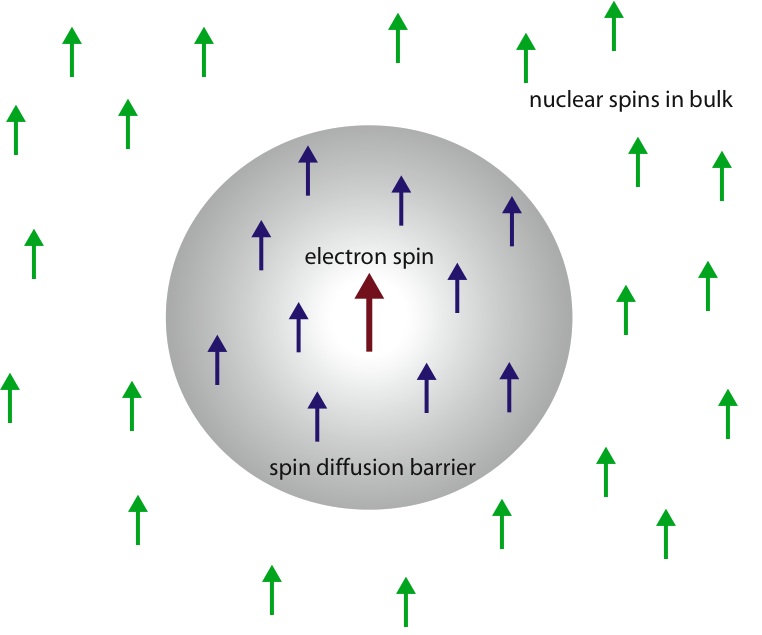}} \caption{\label{diffusionbarrier} Schematic illustration of the spin diffusion barrier around the electron spin.  Nuclear spin diffusion within the barrier is suppressed due to the large difference in Zeeman energies between the spins. }
\end{figure}

While much attention is paid to improving DNP enhancement local to the electron or defect spin, usually by incorporating the appropriate electron spins in the sample, the rate-limiting step in efficiently polarizing bulk samples is frequently the spin transport from the defect sites to the bulk.
The transport  of polarization through the sample occurs via nuclear spin diffusion, which is mediated by energy-conserving flip-flop (XY) interactions.  However, the nuclear spins in an inner core around the impurity experience a large local field gradient due to the impurity spin, and as a consequence have significantly different effective Zeeman energies (Figure \ref{diffusionbarrier}).  
This energy difference suppresses the XY interaction, creating a ``spin-diffusion barrier" around the impurity \cite{Bloembergen-1949}.  In order for the increased polarization to be transferred to the bulk, the polarization needs to be transported across this barrier.  
It has been suggested that spin transport across the barrier could be facilitated by the presence of another energy reservoir, such as the  electron spin dipolar reservoir \cite{Wolf-1973} or the nuclear spin dipolar reservoir \cite{Redfield-1968,Horvitz-1971,Genack-1973,Genack-1975}. The additional reservoir is then able to take up the energy difference between the spins.  

Cooling of the nuclear dipolar reservoir via this mechanism has been observed in optical pumping experiments on InP \cite{Michal-1998}, and GaAs \cite{Patel-2004}.  Here we report on the results of experiments in which the bulk nuclear dipolar reservoir is cooled to a much lower spin temperature than the nuclear Zeeman reservoir, following DNP microwave irradiation.  

Our experiments were performed on a frozen solution containing 40 mM of the nitroxide 4-amino-TEMPO dissolved in a 40:60 water/glycerol mixture.  The experiments were performed in a 2.35 T ($B_0$) superconducting NMR magnet.  An Oxford NMR Spectrostat was operated in single-shot mode to cool the sample down to 1.4 K.   The electron spin frequency was 66 GHz (corresponding to g $\approx$ 2), and the proton Larmor frequency was 100 MHz.  The microwave source was a 90 mW Gunn diode source (Millitech). The NMR spectrometer used was a Bruker Avance system with a home-built probe, containing a horn antenna for the microwaves, and a solenoidal RF coil. Details of the probe have been described elsewhere \cite{Cho-2007}.  In the experiments here, the sample as well as the RF and microwave electronics were in direct contact with the liquid helium bath. 

\begin{figure} \scalebox{0.38}{\includegraphics{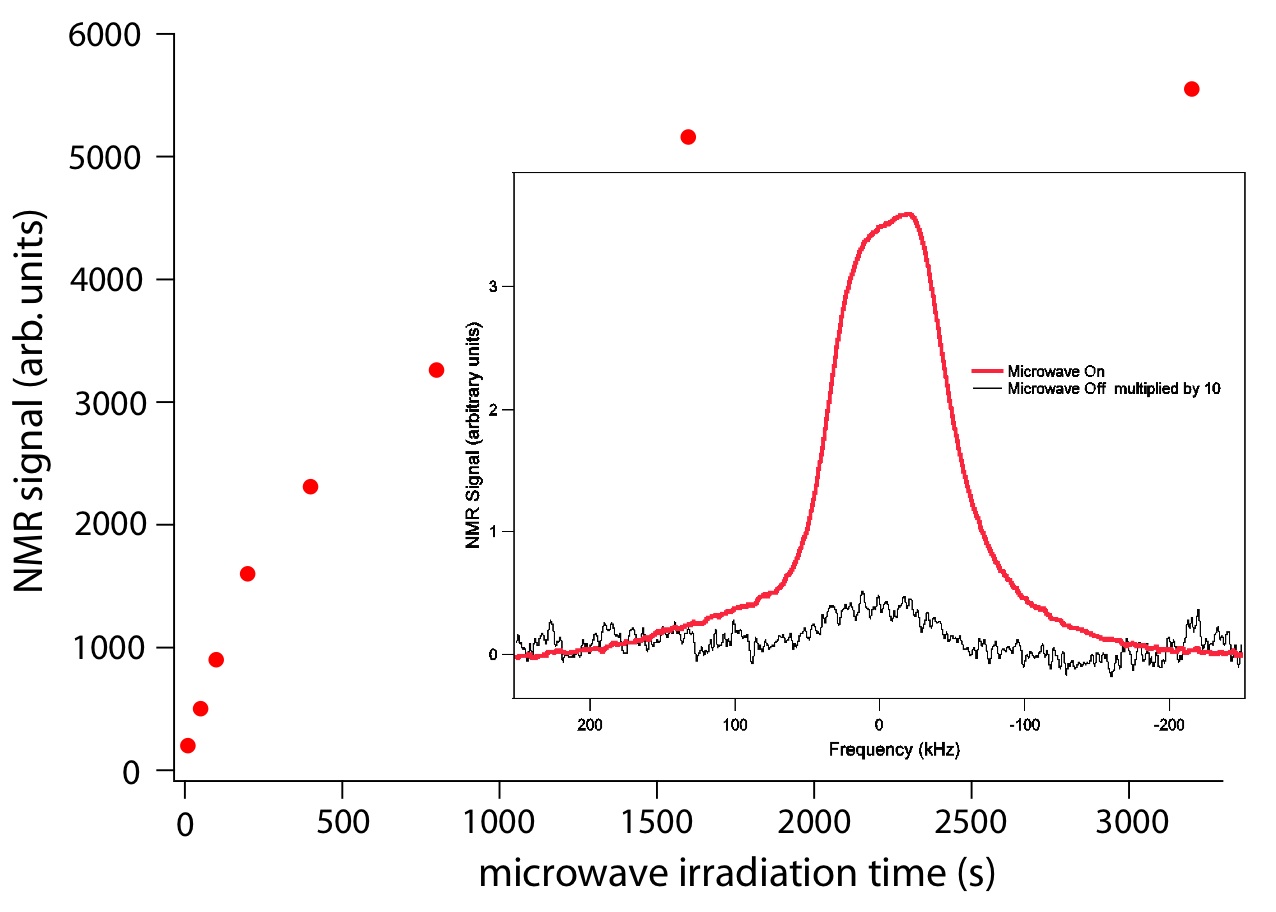}} \caption{\label{DNP} Growth of the NMR signal as a function of the DNP microwave irradiation time.  The inset shows the NMR
spectra demonstrating $^1$H signal enhancement of about 70 following microwave irradiation for 3200 s.}
\end{figure}

Figure~\ref{DNP} shows the growth of the NMR signal as a function of the microwave irradiation time.  The enhancement appears to saturate at around one hour.  The inset shows the NMR spectra  following a $\pi/2$ pulse (= $3 \mu$s) after 3200 s of microwave irradiation and without microwave irradiation.  The observed DNP enhancement is 70, yielding a proton polarization of approximately 12 \%.  The DNP mechanism here is thermal mixing.  Though the 25-35 MHz electron dipolar coupling in this sample is less than the proton Zeeman frequency of 100 MHz, both thermal mixing processes described above should play a role.   

\begin{figure}
\scalebox{0.4}{\includegraphics{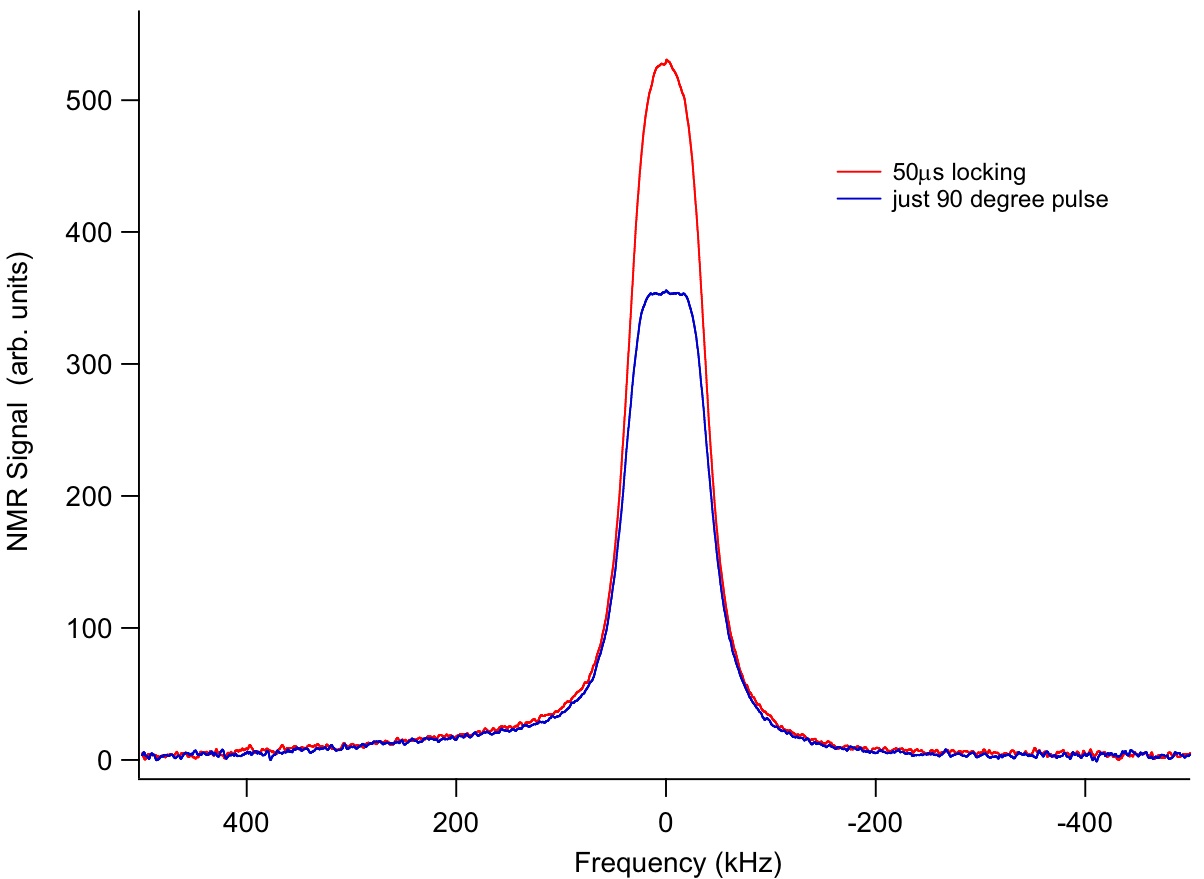}} \caption{\label{DNP-dip} NMR spectra showing a signal enhancement of 1.5 following on-resonance spin-locking with an 83 kHz RF field for 50 $\mu$s.  The microwave irraditation time used was 800 s, yielding a DNP enhancement of 40.}
\end{figure}

Figure~\ref{DNP-dip} compares the signals obtained after microwave irradiation for 800 s, following a $\pi/2$ pulse, and following the $\pi/2$ pulse with a 50 $\mu$s spin locking pulse ($\gamma B_1/2\pi = 83$ kHz).  At 800 s the observed DNP signal enhancement is 40.  As the RF field is comparable to the strength of the local proton dipolar field in the sample ($\gamma B_{\textrm{loc}}/2 \pi \approx$ 85 kHz, estimated from the proton linewidth), it induces mixing between the nuclear Zeeman and nuclear dipolar reservoirs \cite{Hartmann-1962,Anderson-1962,Lurie-1964,Jeener-1965,McArthur-1969,Ramanathan-1996a,Ramanathan-1997}.  The signal increase obtained following this mixing indicates that, following microwave irradiation, the bulk dipolar reservoir had a lower effective spin temperature than the Zeeman reservoir.  

In a strong magnetic field, the Hamiltonian of the bulk nuclear spin system is given by
\begin{eqnarray}
\mathcal{H} & = & \mathcal{H}_Z + \mathcal{H}_D \nonumber \\ & = &  \hbar\omega\sum_iI_z^i + \hbar\sum_{i<j} d_{ij}\left(2I_z^iI_z^j - I_x^iI_x^j - I_y^iI_y^j\right)
\end{eqnarray}
where $\omega = \gamma B_0$, and $d_{ij} = \gamma^2\hbar (1-3\cos^2\vartheta_{ij})/(2r_{ij}^3)$, $r_{ij}$ is the distance between spins $i$ and $j$, and $\vartheta_{ij}$ is the angle between the internuclear vector and the external field.  The Zeeman and dipolar energies are independently conserved.  
We can therefore define independent spin temperatures for each reservoir as well as spin diffusion and spin-lattice relaxation rates.
Note that this is not true at the site of the electron spin defect.
If the spin temperatures of the Zeeman and dipolar reservoir are $\theta_Z$ and $\theta_D$ respectively, the density matrix of the system is given by 
\begin{equation}
\rho = \frac{\exp\left(-k\mathcal{H}_Z/\theta_Z -k\mathcal{H}_D/\theta_D\right)}{Tr\left\{\exp\left(-k\mathcal{H}_Z/\theta_Z -k\mathcal{H}_D/\theta_D \right)\right\}} \: \: .
\end{equation}
We can use a simple thermodynamic argument to estimate the spin temperature of the dipolar reservoir.

After 800 s of microwave irradiation the observed signal enhancement is 40, and the spin temperature of the Zeeman reservoir is $\theta_Z^l \approx 1.4/40 \approx 35$ mK.  
The heat capacities of the Zeeman and dipolar reservoirs are given by $C_Z = C\cdot B_0^2$ and $C_D = C \cdot B_{\textrm{loc}}^2$, where C is the Curie constant, and their energies are $E_Z = -C_Z / k\theta_Z$ and $E_D = - C_D / k\theta_D$. During the spin-locking pulse the magnetization is locked along the applied field ($B_1$), and the heat capacity of the rotating-frame Zeeman reservoir is $C_Z^r = C\cdot B_1^2$.  However, since the energy of the reservoir has not changed, the effective Zeeman spin temperature in the rotating frame is 
$\theta_Z^r = \theta_Z^l (B_1/B_0) \approx 30$ $\mu$K.  

If the magnitude of the spin-locking field is comparable to the strength of the local dipolar field, the two reservoirs are no longer isolated from each other.  During this mixing process the spin temperatures of the Zeeman and dipolar systems equilibrate, but the total energy is conserved.  
Equating the energy of the system before and after the mixing process we get
\begin{equation}
\frac{B_1^2}{\theta_Z^r} + \frac{B_{\textrm{loc}}^2}{\theta_D} = \frac{B_1^2 + B_{\textrm{loc}}^2}{\theta_{eq}^r} \: \: .
\label{eq:thermalmix}
\end{equation}
where $\theta_{eq}^r$ is the final equilibrium temperature after the mixing.  For $B_1 = 83$ kHz, we observe an enhancement of 1.5, which yields a dipolar temperature of 15.5 $\mu$K.    The final laboratory frame Zeeman temperature is 23 mK.  

If we use a long DNP irradiation (3200s sec as shown in Figure 1), and replace the $\pi/2$ pulse and spin locking step with an adiabatic transfer using an adiabatic remagnetization in the rotating frame (ARRF) \cite{Slichter-1961,Slichter-1990}, we have
\begin{equation}
\frac{B_0}{\theta_Z^l}+ \frac{B_{\textrm{loc}}}{\theta_D} = \frac{B_0}{\theta_Z^f}
\label{eq:adiabatic}
\end{equation}
yielding a final Zeeman temperature $\theta_Z^f= $ 9.5 mK.
If the electron spin polarization were transferred directly to the nuclear Zeeman reservoir, we would expect a nuclear spin temperature that is lower than the lattice temperature by a factor $\gamma_e / \gamma_n$ or about 2 mK.

\begin{figure}
\scalebox{0.5}{\includegraphics{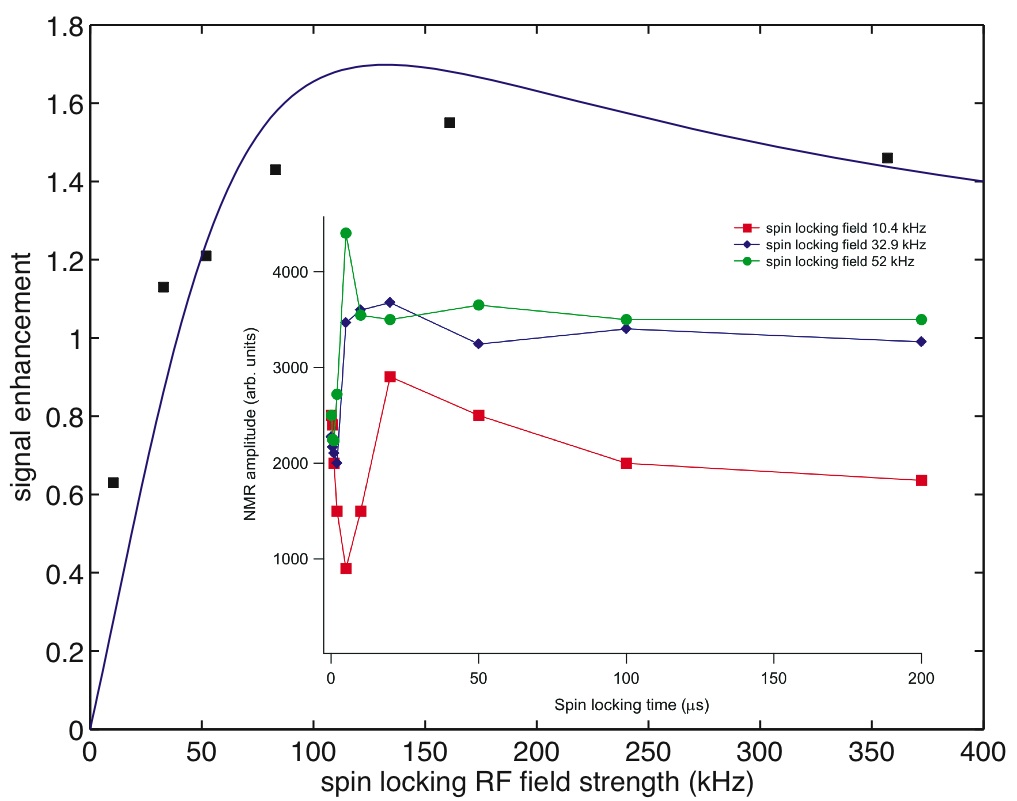}} 
\caption{The additional enhancement measured at a spin-locking time of 200 $\mu$s as a function of the amplitude of the spin locking field.  The solid line is the best fit to Equation~\ref{eq:thermalmix}, yielding a dipolar spin temperature of 48 $\mu$ K.  The inset shows the NMR signal intensity as a function of the spin-locking time for RF field strengths of 10.4, 32.9 and 52 kHz.  \label{fig:oscillations}}
\end{figure}

In order to explore the dynamics of the mixing process, we repeated the experiment for a variety of spin-locking field strengths and times.  The microwave irradiation time used in these experiments was 100 s, which yielded a signal enhancement of 11 (or a 123 mK Zeeman spin temperature).  Figure~\ref{fig:oscillations} shows the signal measured enhancement at a spin-locking time of 200 $\mu$s for different RF field strengths.  The solid line is the best fit to Equation~\ref{eq:thermalmix}, yielding a dipolar spin temperature of 48 $\mu$K. It can be seen that the thermodynamic model fails at lower field strengths as the spins are no longer locked along the RF.

The inset shows the amplitude of the observed signal as a function of the spin-locking time for different RF field strengths.   
There is a transient oscillatory exchange of magnetization between the Zeeman and dipolar reservoirs at short times after the spin-locking field is turned on. These oscillations occur at the Rabi frequency of the spins, and are the rotating frame analog of Strombotne-Hahn oscillations \cite{Strombotne-1964,McArthur-1969,Demco-1975,Ramanathan-1996a}.  The transient oscillations are rapidly damped by the strong proton dipolar couplings in this system.    The 200 $\mu$s time point was chosen to ensure that the the system was in quasi-equilibrium, such that the transient oscillations had damped away and that $T_{1\rho}$ effects were minimal. 

It is worthwhile to examine the possible origins of the this dipolar ordering following DNP.  Genack and Redfied pointed out that the transport of polarization through the magnetic field gradients around the impurity can drive the nuclear dipolar reservoir far out of equilibrium, as it supplies the excess energy required for the transport \cite{Genack-1973,Genack-1975,Jeener}. They derived a set of coupled differential equations to describe the mixing between the Zeeman and dipolar systems in a magnetic field gradient.  We believe that this is the mechanism responsible for the cooling of the dipolar system observed here.

It is also possible that the nuclear spin dipolar reservoir could directly be cooled during microwave irradiation. Tycko \cite{Tycko-1998} showed that dipolar order between two nuclear spins could be created in the presence of fluctuating hyperfine couplings.  However since the thermal mixing process is itself a thermodynamic mixing between the electron dipolar and nuclear Zeeman reservoirs, it is unlikely to result in a direct cooling of the nuclear dipolar reservoir.

We recently measured the spin diffusion rate for dipolar order to be significantly faster than that for Zeeman order \cite{Zhang-1998a,Boutis-2004}.  The experimentally measured speed-up of a factor of 4 (depending on crystal orientation), was greater than that predicted theoretically.  This rapid spin diffusion implies that the cooling of the dipolar reservoir that is produced locally at the spin diffusion barrier, is quickly transferred to the bulk. 

\begin{figure}
\scalebox{0.5}{\includegraphics{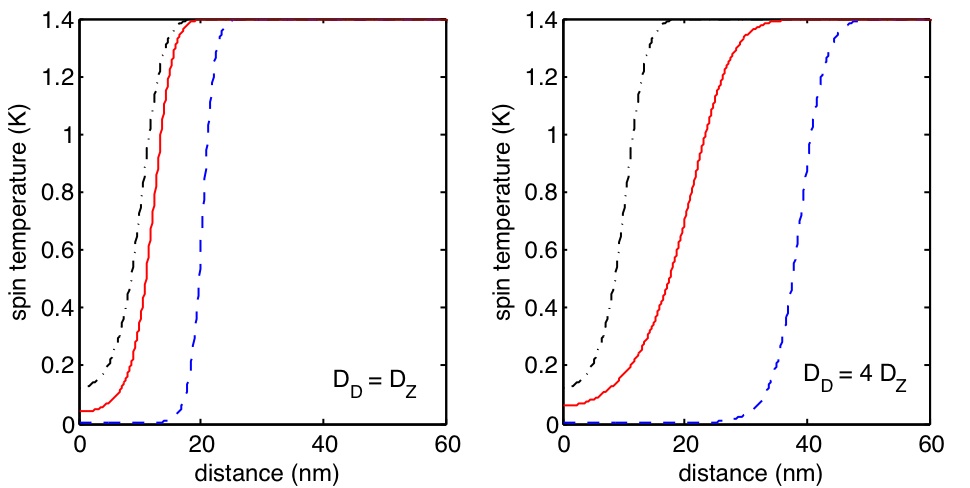}} \caption{Zeeman (black dash-dot line) and dipolar (blue dashed line) spin temperatures  following 1 second of spin diffusion, obtained from simulations where (a) $D_D = D_Z$ and (b) $D_D = 4 D_Z$.  The inital spin temperatures at the left edge are 10 $\mu$K and 20 mK for the dipolar and Zeeman reservoirs respectively.  The solid red line indicates the final spin temperature following adiabatic transfer using Equation~\ref{eq:adiabatic}  \label{fig:sims}}
\end{figure}

The ability to increase the Zeeman magnetization via contact with the dipolar reservoir is exciting, as it should be possible to polarize a sample more rapidly by repeatedly cooling the dipolar reservoir and transferring this polarization to the Zeeman reservoir.  Note that this transfer of order occurs in the bulk crystal, not just locally to the defect sites.  When the weak spin-locking field is applied the cross polarization between the Zeeman and dipolar systems occurs on a much faster timescale, since it does not require macroscopic transport of the polarization.  

Figure \ref{fig:sims} shows the change in Zeeman (black dash-dot line) and dipolar (red dashed line) spin temperatures following spin diffusion for 1 second, obtained from simulation.  In both cases the local dipolar temperature at the origin is 10 $\mu$K and the local Zeeman temperature is 20 mK.  In the figure on the left the dipolar diffusion rate is set equal to the Zeeman diffusion rate ($D_Z = 10^{-5} \mu$m$^2$/s), and in the figure on the right the dipolar diffusion rate is four times faster than the Zeeman diffusion rate.   The figure also shows the final spin temperature that would be reached following an adiabatic tranfer of dipolar to Zeeman order using Equation~\ref{eq:adiabatic} (red solid line).  It can be seen that dipolar order is transported rapidly to the bulk, and that this results in a lower combined spin temperature of the two systems following equilibration.  The transport is not limited by the small heat capacity of the dipolar reservoir.

{\bf Acknowledgements}  This work was supported in part by the National Security Agency (NSA) under Army Research Office (ARO) contract number DAAD190310125, the NSF and DARPA DSO.

\bibliography{Bibliography}

\begin{thebibliography}{26}
\expandafter\ifx\csname natexlab\endcsname\relax\def\natexlab#1{#1}\fi
\expandafter\ifx\csname bibnamefont\endcsname\relax
  \def\bibnamefont#1{#1}\fi
\expandafter\ifx\csname bibfnamefont\endcsname\relax
  \def\bibfnamefont#1{#1}\fi
\expandafter\ifx\csname citenamefont\endcsname\relax
  \def\citenamefont#1{#1}\fi
\expandafter\ifx\csname url\endcsname\relax
  \def\url#1{\texttt{#1}}\fi
\expandafter\ifx\csname urlprefix\endcsname\relax\def\urlprefix{URL }\fi
\providecommand{\bibinfo}[2]{#2}
\providecommand{\eprint}[2][]{\url{#2}}

\bibitem[{\citenamefont{Abragam and Goldman}(1982)}]{Abragam-1982}
\bibinfo{author}{\bibfnamefont{A.}~\bibnamefont{Abragam}} \bibnamefont{and}
  \bibinfo{author}{\bibfnamefont{M.}~\bibnamefont{Goldman}},
  \emph{\bibinfo{title}{Nuclear Magnetism: Order and Disorder}}
  (\bibinfo{publisher}{Clarendon Press}, \bibinfo{address}{Oxford},
  \bibinfo{year}{1982}).

\bibitem[{\citenamefont{Bloembergen}(1949)}]{Bloembergen-1949}
\bibinfo{author}{\bibfnamefont{N.}~\bibnamefont{Bloembergen}},
  \bibinfo{journal}{Physica} \textbf{\bibinfo{volume}{15}},
  \bibinfo{pages}{386} (\bibinfo{year}{1949}).

\bibitem[{\citenamefont{Farrar et~al.}(2001)\citenamefont{Farrar, Hall, Gerfen,
  Inati, and Griffin}}]{Farrar-2001}
\bibinfo{author}{\bibfnamefont{C.~T.} \bibnamefont{Farrar}},
  \bibinfo{author}{\bibfnamefont{D.~A.} \bibnamefont{Hall}},
  \bibinfo{author}{\bibfnamefont{G.~J.} \bibnamefont{Gerfen}},
  \bibinfo{author}{\bibfnamefont{S.~J.} \bibnamefont{Inati}}, \bibnamefont{and}
  \bibinfo{author}{\bibfnamefont{R.~G.} \bibnamefont{Griffin}},
  \bibinfo{journal}{J. Chem. Phys.} \textbf{\bibinfo{volume}{114}},
  \bibinfo{pages}{4922} (\bibinfo{year}{2001}).

\bibitem[{\citenamefont{Reynhardt}(2003)}]{Reynhardt-2003b}
\bibinfo{author}{\bibfnamefont{E.~C.} \bibnamefont{Reynhardt}},
  \bibinfo{journal}{Concepts Magn. Reson. A} \textbf{\bibinfo{volume}{19A}},
  \bibinfo{pages}{36} (\bibinfo{year}{2003}).

\bibitem[{\citenamefont{Wolfe}(1973)}]{Wolf-1973}
\bibinfo{author}{\bibfnamefont{J.~P.} \bibnamefont{Wolfe}},
  \bibinfo{journal}{Phys. Rev. Lett.} \textbf{\bibinfo{volume}{31}},
  \bibinfo{pages}{907} (\bibinfo{year}{1973}).

\bibitem[{\citenamefont{Redfield and Yu}(1968)}]{Redfield-1968}
\bibinfo{author}{\bibfnamefont{A.~G.} \bibnamefont{Redfield}} \bibnamefont{and}
  \bibinfo{author}{\bibfnamefont{W.~N.} \bibnamefont{Yu}},
  \bibinfo{journal}{Phys. Rev.} pp. \bibinfo{pages}{443--450}
  (\bibinfo{year}{1968}).

\bibitem[{\citenamefont{Horvitz}(1971)}]{Horvitz-1971}
\bibinfo{author}{\bibfnamefont{E.~P.} \bibnamefont{Horvitz}},
  \bibinfo{journal}{Phys. Rev. B} \textbf{\bibinfo{volume}{3}},
  \bibinfo{pages}{2868} (\bibinfo{year}{1971}).

\bibitem[{\citenamefont{Genack and Redfield}(1973)}]{Genack-1973}
\bibinfo{author}{\bibfnamefont{A.~Z.} \bibnamefont{Genack}} \bibnamefont{and}
  \bibinfo{author}{\bibfnamefont{A.~G.} \bibnamefont{Redfield}},
  \bibinfo{journal}{Phys. Rev. Lett.} \textbf{\bibinfo{volume}{31}},
  \bibinfo{pages}{1204} (\bibinfo{year}{1973}).

\bibitem[{\citenamefont{Genack and Redfield}(1975)}]{Genack-1975}
\bibinfo{author}{\bibfnamefont{A.~Z.} \bibnamefont{Genack}} \bibnamefont{and}
  \bibinfo{author}{\bibfnamefont{A.~G.} \bibnamefont{Redfield}},
  \bibinfo{journal}{Phys. Rev. B} \textbf{\bibinfo{volume}{12}},
  \bibinfo{pages}{78} (\bibinfo{year}{1975}).

\bibitem[{\citenamefont{Michal and Tycko}(1998)}]{Michal-1998}
\bibinfo{author}{\bibfnamefont{C.~A.} \bibnamefont{Michal}} \bibnamefont{and}
  \bibinfo{author}{\bibfnamefont{R.}~\bibnamefont{Tycko}},
  \bibinfo{journal}{Phys. Rev. Lett.} \textbf{\bibinfo{volume}{81}},
  \bibinfo{pages}{3988} (\bibinfo{year}{1998}).

\bibitem[{\citenamefont{Patel and Bowers}(2004)}]{Patel-2004}
\bibinfo{author}{\bibfnamefont{A.}~\bibnamefont{Patel}} \bibnamefont{and}
  \bibinfo{author}{\bibfnamefont{C.~R.} \bibnamefont{Bowers}},
  \bibinfo{journal}{Chem. Phys. Lett.} \textbf{\bibinfo{volume}{397}},
  \bibinfo{pages}{96} (\bibinfo{year}{2004}).

\bibitem[{\citenamefont{Cho et~al.}(2007)\citenamefont{Cho, Baugh, Ryan, Cory,
  and Ramanathan}}]{Cho-2007}
\bibinfo{author}{\bibfnamefont{H.}~\bibnamefont{Cho}},
  \bibinfo{author}{\bibfnamefont{J.}~\bibnamefont{Baugh}},
  \bibinfo{author}{\bibfnamefont{C.~A.} \bibnamefont{Ryan}},
  \bibinfo{author}{\bibfnamefont{D.~G.} \bibnamefont{Cory}}, \bibnamefont{and}
  \bibinfo{author}{\bibfnamefont{C.}~\bibnamefont{Ramanathan}},
  \bibinfo{journal}{J. Magn. Reson.} \textbf{\bibinfo{volume}{187}},
  \bibinfo{pages}{242} (\bibinfo{year}{2007}).

\bibitem[{\citenamefont{Hartmann and Hahn}(1962)}]{Hartmann-1962}
\bibinfo{author}{\bibfnamefont{S.~R.} \bibnamefont{Hartmann}} \bibnamefont{and}
  \bibinfo{author}{\bibfnamefont{E.~L.} \bibnamefont{Hahn}},
  \bibinfo{journal}{Phys. Rev.} \textbf{\bibinfo{volume}{128}},
  \bibinfo{pages}{2042} (\bibinfo{year}{1962}).

\bibitem[{\citenamefont{Anderson and Hartmann}(1962)}]{Anderson-1962}
\bibinfo{author}{\bibfnamefont{A.~G.} \bibnamefont{Anderson}} \bibnamefont{and}
  \bibinfo{author}{\bibfnamefont{S.~R.} \bibnamefont{Hartmann}},
  \bibinfo{journal}{Phys. Rev.} \textbf{\bibinfo{volume}{128}},
  \bibinfo{pages}{2023} (\bibinfo{year}{1962}).

\bibitem[{\citenamefont{Lurie and Slichter}(1964)}]{Lurie-1964}
\bibinfo{author}{\bibfnamefont{F.~M.} \bibnamefont{Lurie}} \bibnamefont{and}
  \bibinfo{author}{\bibfnamefont{C.~P.} \bibnamefont{Slichter}},
  \bibinfo{journal}{Phys. Rev.} \textbf{\bibinfo{volume}{133}},
  \bibinfo{pages}{A1108} (\bibinfo{year}{1964}).

\bibitem[{\citenamefont{Jeener et~al.}(1965)\citenamefont{Jeener, Du~Bois, and
  Broekaert}}]{Jeener-1965}
\bibinfo{author}{\bibfnamefont{J.}~\bibnamefont{Jeener}},
  \bibinfo{author}{\bibfnamefont{R.}~\bibnamefont{Du~Bois}}, \bibnamefont{and}
  \bibinfo{author}{\bibfnamefont{P.}~\bibnamefont{Broekaert}},
  \bibinfo{journal}{Phys. Rev.} \textbf{\bibinfo{volume}{139}},
  \bibinfo{pages}{A1959} (\bibinfo{year}{1965}).

\bibitem[{\citenamefont{McArthur et~al.}(1969)\citenamefont{McArthur, Hahn, and
  Walstedt}}]{McArthur-1969}
\bibinfo{author}{\bibfnamefont{D.~A.} \bibnamefont{McArthur}},
  \bibinfo{author}{\bibfnamefont{E.~L.} \bibnamefont{Hahn}}, \bibnamefont{and}
  \bibinfo{author}{\bibfnamefont{R.~E.} \bibnamefont{Walstedt}},
  \bibinfo{journal}{Phys. Rev.} \textbf{\bibinfo{volume}{188}},
  \bibinfo{pages}{609} (\bibinfo{year}{1969}).

\bibitem[{\citenamefont{Ramanathan et~al.}(1996)\citenamefont{Ramanathan, Wu,
  Pfleiderer, Lizak, Garrido, and Ackerman}}]{Ramanathan-1996a}
\bibinfo{author}{\bibfnamefont{C.}~\bibnamefont{Ramanathan}},
  \bibinfo{author}{\bibfnamefont{Y.}~\bibnamefont{Wu}},
  \bibinfo{author}{\bibfnamefont{B.}~\bibnamefont{Pfleiderer}},
  \bibinfo{author}{\bibfnamefont{M.~J.} \bibnamefont{Lizak}},
  \bibinfo{author}{\bibfnamefont{L.}~\bibnamefont{Garrido}}, \bibnamefont{and}
  \bibinfo{author}{\bibfnamefont{J.~L.} \bibnamefont{Ackerman}},
  \bibinfo{journal}{J. Magn. Reson. A} \textbf{\bibinfo{volume}{121}},
  \bibinfo{pages}{127} (\bibinfo{year}{1996}).

\bibitem[{\citenamefont{Ramanathan and Ackerman}(1997)}]{Ramanathan-1997}
\bibinfo{author}{\bibfnamefont{C.}~\bibnamefont{Ramanathan}} \bibnamefont{and}
  \bibinfo{author}{\bibfnamefont{J.~L.} \bibnamefont{Ackerman}},
  \bibinfo{journal}{J. Magn. Reson.} \textbf{\bibinfo{volume}{127}},
  \bibinfo{pages}{26} (\bibinfo{year}{1997}).

\bibitem[{\citenamefont{Slichter and Holton}(1961)}]{Slichter-1961}
\bibinfo{author}{\bibfnamefont{C.~P.} \bibnamefont{Slichter}} \bibnamefont{and}
  \bibinfo{author}{\bibfnamefont{W.~C.} \bibnamefont{Holton}},
  \bibinfo{journal}{Phys. Rev.} \textbf{\bibinfo{volume}{122}},
  \bibinfo{pages}{1701} (\bibinfo{year}{1961}).

\bibitem[{\citenamefont{Slichter}(1990)}]{Slichter-1990}
\bibinfo{author}{\bibfnamefont{C.~P.} \bibnamefont{Slichter}},
  \emph{\bibinfo{title}{Principles of Magnetic Resonance}}
  (\bibinfo{publisher}{Springer-Verlag}, \bibinfo{address}{Berlin},
  \bibinfo{year}{1990}), \bibinfo{edition}{3rd} ed.

\bibitem[{\citenamefont{Strombotne and Hahn}(1964)}]{Strombotne-1964}
\bibinfo{author}{\bibfnamefont{R.~L.} \bibnamefont{Strombotne}}
  \bibnamefont{and} \bibinfo{author}{\bibfnamefont{E.~L.} \bibnamefont{Hahn}},
  \bibinfo{journal}{Phys. Rev.} \textbf{\bibinfo{volume}{133}},
  \bibinfo{pages}{A1616} (\bibinfo{year}{1964}).

\bibitem[{\citenamefont{Demco et~al.}(1975)\citenamefont{Demco, Tegenfeldt, and
  Waugh}}]{Demco-1975}
\bibinfo{author}{\bibfnamefont{D.~E.} \bibnamefont{Demco}},
  \bibinfo{author}{\bibfnamefont{J.}~\bibnamefont{Tegenfeldt}},
  \bibnamefont{and} \bibinfo{author}{\bibfnamefont{J.~S.} \bibnamefont{Waugh}},
  \bibinfo{journal}{Phys. Rev. B} \textbf{\bibinfo{volume}{11}},
  \bibinfo{pages}{4133} (\bibinfo{year}{1975}).

\bibitem[{\citenamefont{Jeener}()}]{Jeener}
\bibinfo{author}{\bibfnamefont{J.}~\bibnamefont{Jeener}},
  \bibinfo{misc}{personal communication}.
  
\bibitem[{\citenamefont{Tycko}(1998)}]{Tycko-1998}
\bibinfo{author}{\bibfnamefont{R.}~\bibnamefont{Tycko}}, \bibinfo{journal}{Mol.
  Phys.} \textbf{\bibinfo{volume}{95}}, \bibinfo{pages}{1169}
  (\bibinfo{year}{1998}).

\bibitem[{\citenamefont{Zhang and Cory}(1998)}]{Zhang-1998a}
\bibinfo{author}{\bibfnamefont{W.}~\bibnamefont{Zhang}} \bibnamefont{and}
  \bibinfo{author}{\bibfnamefont{D.~G.} \bibnamefont{Cory}},
  \bibinfo{journal}{Phys. Rev. Lett.} \textbf{\bibinfo{volume}{80}},
  \bibinfo{pages}{1324} (\bibinfo{year}{1998}).

\bibitem[{\citenamefont{Boutis et~al.}(2004)\citenamefont{Boutis, Greenbaum,
  Cho, Cory, and Ramanathan}}]{Boutis-2004}
\bibinfo{author}{\bibfnamefont{G.~S.} \bibnamefont{Boutis}},
  \bibinfo{author}{\bibfnamefont{D.}~\bibnamefont{Greenbaum}},
  \bibinfo{author}{\bibfnamefont{H.}~\bibnamefont{Cho}},
  \bibinfo{author}{\bibfnamefont{D.~G.} \bibnamefont{Cory}}, \bibnamefont{and}
  \bibinfo{author}{\bibfnamefont{C.}~\bibnamefont{Ramanathan}},
  \bibinfo{journal}{Phys. Rev. Lett.} \textbf{\bibinfo{volume}{92}},
  \bibinfo{pages}{137201} (\bibinfo{year}{2004}).

\end{thebibliography}

\end{document}